\documentclass[twocolumn,prb]{revtex4-2}
\usepackage[usenames,dvipsnames]{color}

\usepackage{comment}
\usepackage{graphicx}
\usepackage{inputenc}
\usepackage{lineno}
\usepackage{color}
\usepackage{hyperref}
\usepackage{amsmath}
\usepackage{braket}
\usepackage{caption}
\usepackage{bm}
\usepackage{subcaption}

\usepackage[]{todonotes}

\begin{document}

\title{Emergent nearest-neighbor attraction in the fully renormalized interactions of the single-band repulsive Hubbard model at weak coupling.}
\author{Daria Gazizova}
\affiliation{Department of Physics and Physical Oceanography, Memorial University of Newfoundland, St. John's, Newfoundland \& Labrador, Canada A1B 3X7} 
\author{J. P. F. LeBlanc}
\email{jleblanc@mun.ca}
\affiliation{Department of Physics and Physical Oceanography, Memorial University of Newfoundland, St. John's, Newfoundland \& Labrador, Canada A1B 3X7}

\date{\today}
\begin{abstract}
 We compute the perturbative expansion for the effective interaction $W$ of the half-filled 2-dimensional Hubbard model.  We derive extensions of standard RPA resummations that include arbitrarily high order contributions in the $W_{\uparrow\uparrow}$ and $W_{\uparrow\downarrow}$ basis. 
 Using algorithmic tools we explore the static $Q$-dependent interaction as well as the same-time quantity both in momentum- and real-space.  We emphasize the absence of screening in the Hubbard interaction where we find an enhanced repulsive local $W_{\uparrow\downarrow}$ with a non-zero attractive $W_{\uparrow\uparrow}$.  Finally, starting from only a locally repulsive bare interaction find an emergent non-local nearest-neighbor attraction for low temperatures at sufficiently large values of $U/t$ which may be key to understanding pairing processes in the model.
\end{abstract}

\maketitle

\section{Introduction}
The Hubbard interaction has been widely studied as a benchmarking tool for the development of methods and algorithms for strongly correlated systems.\cite{benchmarks,footprints,mingpu:2022,esslinger,jarrell:2001:2,kozik:2010,Schaefer:2016,kaufmann:2021}
The model on the 2D square lattice is particularly interesting giving rise to a plethora of phases that are reminiscent of the high-temperature cuprates\cite{jarrell:2001,gull:2013,Maier00,Macridin:2006} as well as more exotic phases such as pseudo-gap, pair-density wave or stripe orders.\cite{wietek:2021,white:2003,Huang:2018,zheng:2017,corboz:stripes} The 2D model itself remains, according to some,\cite{arovas:2022} not well understood. In particular the T=0 phase diagram remains a topic of controversy with evidence both for and against the superconducting phase giving way to stripe ordering.\cite{qin:2020,gong:2021} 
This variety of phases is surprising when one considers that the Hubbard interaction is purely local in real-space and provides only a uniform repulsion in momentum-space.  
  At finite temperatures there has been substantial progress  for the weakly coupled Hubbard model, where a variety of numerical methods are able to agree on the prevalence of $Q=(\pi,\pi)$ spin excitations, the amplitude of the spin-correlation length, as well as the onset of a metal-to-insulator crossover and pseudogap behaviours.\cite{footprints,benchmarks} 

Despite agreement for some observables at weak coupling, there remain fundamental questions about the 2D Hubbard model that do not involve phases or correlation lengths. It is not known how the simple momentum independent $U$ gives effectively non-local interactions nor what the structure of those interactions might be.  Logically, the effective interaction must be the driving force behind any phases that might arise in the model.  In this work we will address this key deficiency in our knowledge by computing the effective renormalized interaction between two spins, $W_{\sigma \sigma^\prime}$, while avoiding controversial aspects of the model with regards to specific phases. To do this we employ state of the art algorithms for symbolic integration of Feynman perturbative expansions.\cite{AMI,taheri:2020,GIT}  These expansions are limited in the range of interaction strength accessible, but have the key advantage that they can be evaluated for infinite systems.  Hence, where these expansions can be converged the results are exact and in the thermodynamic limit.  In addition, we derive expressions for infinitely resummed diagrammatic series from which we can extract the effective interactions in any basis of momentum/real-space or imaginary-time/frequency.

Our results demonstrate that the effective interaction in the Hubbard model is enhanced and not screened by higher order contributions to the interaction with dominant contribution in the $(\pi,\pi)$ region of momentum space.  In addition we find an attractive same-spin interaction that emerges from the locally repulsive term of the Hamiltonian.  In certain parameter ranges we find that the effective interaction becomes spatially oscillatory giving rise to both attractive and repulsive domains with single-lattice size scale. Finally, we comment on the observed scaling behavior of the effective interaction with $U/t$ and temperature.

\section{Models and Methods}
\subsection{Hubbard Hamiltonian}
\label{main:Hubbard}
We study the single-band Hubbard Hamiltonian on a 2D square lattice\cite{benchmarks},
\begin{eqnarray}\label{E:Hubbard}
H = \sum_{ ij \sigma} t_{ij}c_{i\sigma}^\dagger c_{j\sigma} + U\sum_{i} n_{i\uparrow} n_{i\downarrow},
\end{eqnarray}
where $t_{ij}$ is the hopping amplitude, $c_{i\sigma}^{(\dagger)}$ ($c_{i\sigma}$) is the creation (annihilation) operator at site $i$, $\sigma \in \{\uparrow,\downarrow\}$ is the spin, $U$ is the onsite Hubbard interaction, $n_{i\sigma} = c_{i\sigma}^{\dagger}c_{i\sigma}$ is the number operator.  We restrict the sum over sites to nearest and next-nearest neighbors for a 2D square lattice, resulting in the free particle energy 
\begin{eqnarray}
\nonumber\epsilon(\textbf k)=-2t[\cos(k_x)+\cos(k_y)]-\mu,
\end{eqnarray} 
where $\mu$ is the chemical potential, and $t$ is the nearest-neighbor hopping amplitude. Throughout we work with energies in units of the hopping, $t=1$.  We absorb the Hartree shift and restrict our discussion to the $\mu=0$ half-filled case.

\begin{figure}
\begin{minipage}{1\linewidth}
\center{\includegraphics[width=\linewidth]{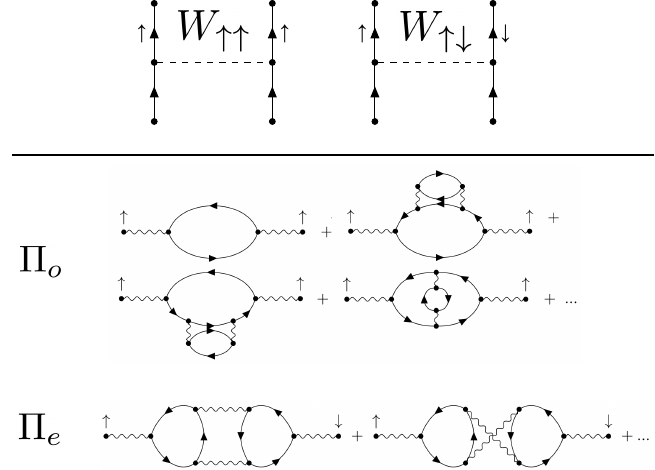}} \\
\end{minipage}
\hfill
\caption{Top: Dashed lines depicting the same- and opposite-spin effective interactions.  Bottom: A subset of Feynman diagrams for the odd polarization $\Pi_o$ and even polarization $\Pi_e$
corrections truncated at second order. Solid lines represent fermionic propagators with onsite U/t interaction depicted as wavy lines.  Bare Hubbard interactions are allowed only between opposite spins, therefore left and right part for $\Pi_{o/e}$ connects same/opposite spins.}
\label{fig:diagrams}
\end{figure}

\subsection{Algorithmic Matsubara Integration}
The algorithmic Matsubara integration (AMI) method introduced in Ref.~\onlinecite{AMI}, provides a versatile approach for analytically evaluating temporal integrals in Feynman diagram expansions. In essence, AMI employs the residue theorem to construct the analytic solution for high-dimensional integrands composed of bare Green's functions. While performing the Matsubara summations themselves is not conceptually difficult and is covered in various textbook exercises, the challenge lies in the exponential growth of the number of analytic terms as the diagram order increases.

By utilizing the existing AMI library\cite{libami}, the AMI result is stored in three nested arrays: Signs/prefactors ($S$), complex poles ($P$), and Green's functions ($R$). These three objects require minimal storage space and enable the construction of the symbolic analytic expression through elementary algebraic operations \cite{AMI}.
The beauty of this approach is that the resulting expression remains analytic in external variables and allows for true analytic continuation of the external Matsubara frequency, $i\nu_n \to \nu +i0^+$, while also being an explicit function of temperature ($T$). Alternatively, the external frequency can also be summed providing access to the same-time object at $\tau=0$. Furthermore, the AMI procedure typically needs to be performed only once for a given graph topology and remains valid for any choice of dispersion in any dimensionality and it can be applied to model systems for a wide range of Feynman diagrammatic expansions \cite{leblanc:2022,taheri:2020,GIT,burke:2023,farid:2023}.

\subsection{Fully Screened Interaction}


Screened interactions play a pivotal role in material calculations, most predominantly in the use of the so-called $GW$-approximation for the single particle self energy.  The self-energy is paramount when comparing energy bands from density functional theory to experimental spectra since it provides the widths of the peaks in the density of states or spectral function.
It is common for $W$ to be approximated via an RPA expansion.  This is done not because the RPA expansion is a good approximation but purely that the analytic expression for the bare bubble - the Lindhard function - is easily derivable and can be evaluated in real frequencies for virtually any system.   Using algorithmic Matsubara integration (AMI) we can compute any diagram in real frequencies and this removes the necessity of studying the RPA-approximation. 

There is, however, merit to the factorization approach of Dyson-like expansions based on bare diagrams.\cite{kozik2015nonexistence, mcniven:2021}  In the case of the Hubbard interaction there is a peculiarity that the bare expansion should only include diagrams with interactions between opposing spins.  There is therefore a natural basis for separating the effective interaction, illustrated in Fig.~\ref{fig:diagrams}. 
We call the effective interaction between opposite spins $W_{\uparrow \downarrow}$ and between same spins $W_{\uparrow \uparrow}$.\cite{Gukelberger:2015}  To proceed we separate the one-particle(bose) irreducible diagrams into those that have either an  odd(o) or even(e) number of bubbles along the principle chain - examples shown in Fig.~\ref{fig:diagrams} -  and we call these polarization diagram sets $\Pi_{o}$ and $\Pi_{e}$ respectively.  We note that these are not new objects and elsewhere are defined to be $\Pi_{\uparrow\uparrow}$ and $\Pi_{\uparrow \downarrow}$ but we find the even/odd description to be more intuitive/instructive.\cite{Rohringer12}  
The full expressions for $W_{\uparrow\uparrow}$ and $W_{\uparrow\downarrow}$ are all chains of reducible combinations that maintain overall odd or even character respectively.  This leads to compact expressions for the infinite resummation in each case in the form of a combinatorics problem.

\begin{table}
    \centering
    \begin{tabular}{|c|c|c|c|c|}
    \hline
        Order & $\Pi_o$ & $\Pi_e$  & $W_{\uparrow\uparrow}$ & $W_{\uparrow\downarrow}$ \\ \hline \hline
        0 & 1  & 0 & 1 & 0\\
         1& 0& 0& 0 & 1 \\
         2 & 3 & 2 & 4 & 2 \\
         3 & 8 & 6&  12& 13 \\
         4 & 65 & 52 &  87 & 74  \\
    \hline
    \end{tabular}
    \caption{\label{tab:nterms}Number of diagrams for the Hubbard interaction in the $\Pi_o$ and $\Pi_e$ expansions (irreducible diagrams) at each order as well as for the full expansions of $W$ including reducible and irreducible diagrams.}   
\end{table} 

In the case of $W_{\uparrow\uparrow}$ one has the expansion
\begin{align}
    W_{\uparrow \uparrow} &= - U \Pi_o U + U \Pi_o U \Pi_e U 
    \\ &+ U \Pi_e U  \Pi_o U - U\Pi_o U\Pi_o U\Pi_o U 
    + ... \\  &= U \sum_{i=0}^\infty \sum_{m=0}^\infty \begin{pmatrix}
2i + 1 +m\\
m
\end{pmatrix}
(-1)^{m+1}
(U\Pi_o)^{2i +1} (U\Pi_e)^m  
\end{align}
which can be replaced with
\begin{equation}\label{eq:wuu}
    W_{\uparrow \uparrow} / U = \frac{-\Pi_o U }{(1+U\Pi_e )^2 - (U\Pi_o)^2}
\end{equation}
which is valid so long as the denominator remains positive.

Similarly, in the case of $W_{\uparrow\downarrow}$ we obtain 
\begin{align}
    W_{\uparrow \downarrow} &=  U - U \Pi_e U + U \Pi_e U \Pi_e U 
    \\ &+ U \Pi_o U  \Pi_o U 
    - U\Pi_o U\Pi_o U\Pi_e U
    - ... \\  &=  U\sum_{i=0}^\infty \sum_{m=0}^\infty \begin{pmatrix}
2i+m\\
m
\end{pmatrix}
 (-1)^{m}(U\Pi_o)^{2i} (U\Pi_e)^m 
\end{align}
resulting in
\begin{equation}\label{eq:wud}
    W_{\uparrow \downarrow} / U = \frac{1 + \Pi_e U }{(1+U\Pi_e )^2 - (U\Pi_o)^2}
\end{equation}
with the same constraint that the denominator be greater than zero.

Equations (\ref{eq:wuu}) and (\ref{eq:wud}) are therefore extensions of RPA for arbitrary truncation of the $\Pi_o$ and $\Pi_e$ expansions.  
Immediately we see some scaling behaviour in the weak-interaction limit.  In the case of Eq.~\ref{eq:wuu} we expect to find $W_{\uparrow\uparrow}/U \propto -\Pi_o U\approx -aU+ \mathcal{O}{({U^2})}$ since the lowest order diagram in $\Pi_o$ is of order $U^0$.  Similarly,  $W_{\uparrow\downarrow}/U \propto 1+\Pi_o U \Pi_o U - \Pi_e U \approx a+bU^2 - cU^3+\mathcal{O}{({U^4})}$.
One can also recover typical RPA expressions by summing equations (\ref{eq:wuu}) and (\ref{eq:wud}) with appropriate replacement of $\Pi_o$ with bare bubble diagram and as $\Pi_e$ approaches zero.
While the weak-interaction scaling between RPA and our full expansions should not change drastically, the normal RPA expansion has a divergence when the denominator approaches zero beyond which it is invalid.  In the case of susceptibilities this divergence is often viewed as a second-order phase transition, but in fact it is simply an artifact of the truncated expansion.  We see in our Eqs.(\ref{eq:wuu}) and (\ref{eq:wud}) that the additional $\Pi_e$ diagrams will actually prevent the divergence from occurring.  

In practice one cannot compute $\Pi_o$ and $\Pi_e$ exactly and instead might compute a truncation of each expansion.  We truncate $\Pi_o$ and $\Pi_e$ at orders $n$ and $m$ respectively and $W_{\sigma \sigma^\prime}$ must therefore depend on this truncation.  When necessary we extend our notation $W_{\sigma \sigma^\prime} \to W_{\sigma \sigma^\prime}^{(n,m)}$ to mark the truncation orders used.  Typically third or fourth order truncation in each diagram series is tractable where the total number of diagrams is $137$ (see Table~\ref{tab:nterms}) which are then infinitely resummed via equations (\ref{eq:wuu}) and (\ref{eq:wud}).
Alternatively, one can compute the $W_{\sigma \sigma^\prime}$ series directly including both reducible and irreducible diagrams up to a single truncation order which we will denote $W_{\sigma \sigma^\prime}^{(l)}$ for truncation at $l$th order.  The reliability of the resummation scheme can therefore be determined through comparison to the direct truncated expansion. 

This separation of $W_{\uparrow\uparrow}$ and $W_{\uparrow\downarrow}$ is distinct from typical screening of a coulomb interaction.  For a spin independent interaction the full effective interaction would just be the sum of the two series.  Since the sign of $\Pi_e$ and $\Pi_o$ are typically different due to the odd/even number of fermionic loops the sum is expected to result in a suppression or screening of the overall interaction.  We will see that this does not happen for the spin-dependent Hubbard interaction where $W_{\uparrow\uparrow}$ and $W_{\uparrow\downarrow}$ don't mix and as a result higher order contributions do not screen the interaction but rather act to enhance it. 

In what follows we compute $\Pi_o(Q,i\Omega_n)$ and $\Pi_e(Q,i\Omega_n)$ where we will study first the static case of $i\Omega_n=0$.  Subsequently we will construct the same-time interaction via the $\tau=0$ Fourier transform 
\begin{equation}\label{eq:sametime}
    W_{\uparrow\uparrow/\uparrow\downarrow}(Q,\tau=0)=\sum\limits_{i\Omega_n}W_{\uparrow\uparrow/\uparrow\downarrow}(Q,i\Omega_n)
\end{equation}
where we again will make use of AMI to exactly sum the infinite set of external frequencies and provide analytic expressions for the same-time diagrams. Finally, we can Fourier transform to real-space and study the effective spatially dependent same-time interaction, $W_{\sigma \sigma^\prime}(r,\tau=0)$.  We will restrict our discussion to the vectors $r=(0,0),(0,1),$ and $(1,1)$ for local, nearest-neighbour and next-nearest-neighbour effective interactions.

\begin{figure}
\centering
\includegraphics[width=1\linewidth]{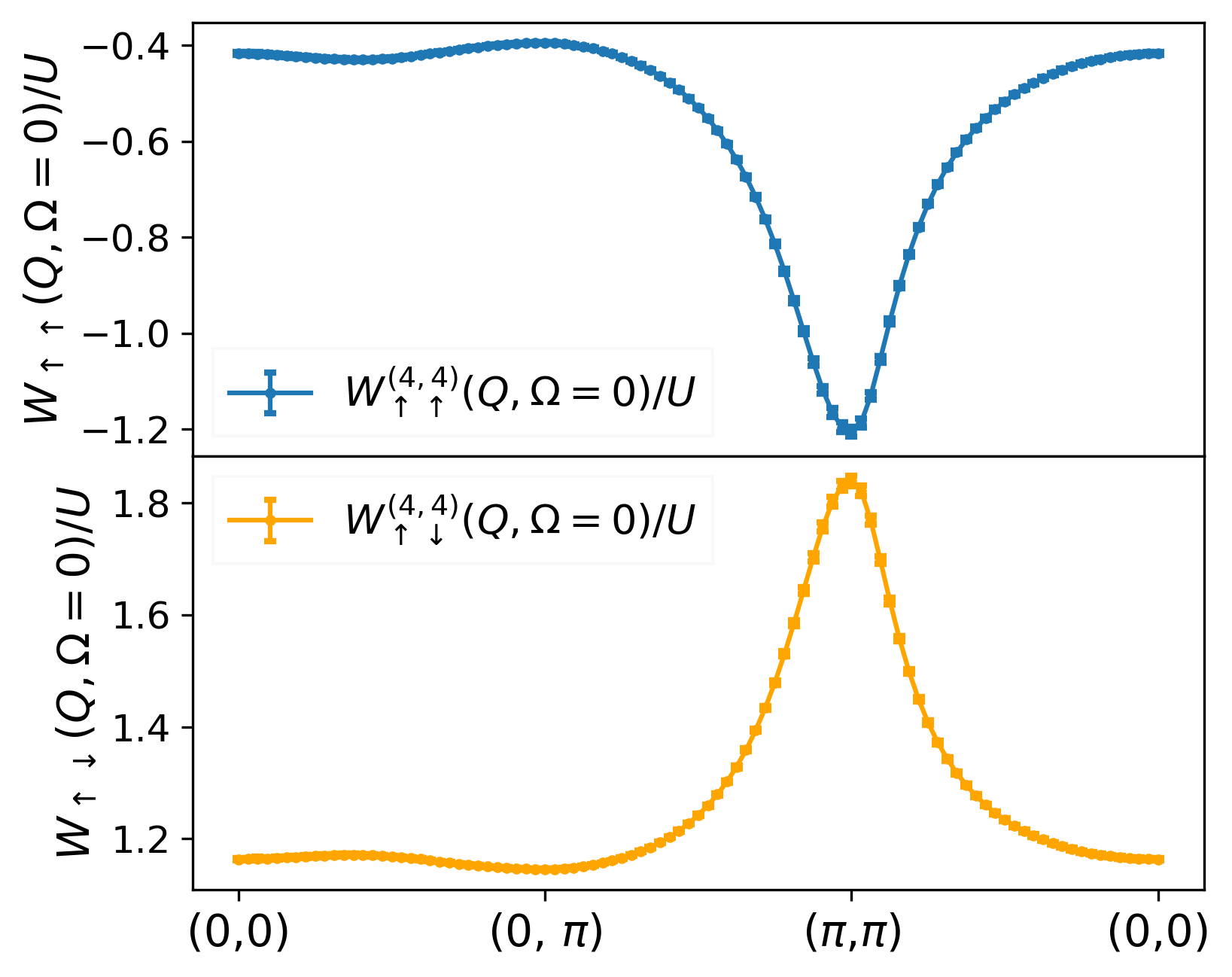}
\caption{\label{fig:static}Static interactions throughout the Brillouin zone for $W_{\sigma\sigma'}^{(4, 4)} /U $ at $\beta t=3$, $U/t=2$ and $\mu=0$. }
\end{figure}

\begin{figure}
\centering
\includegraphics[width=1\linewidth]{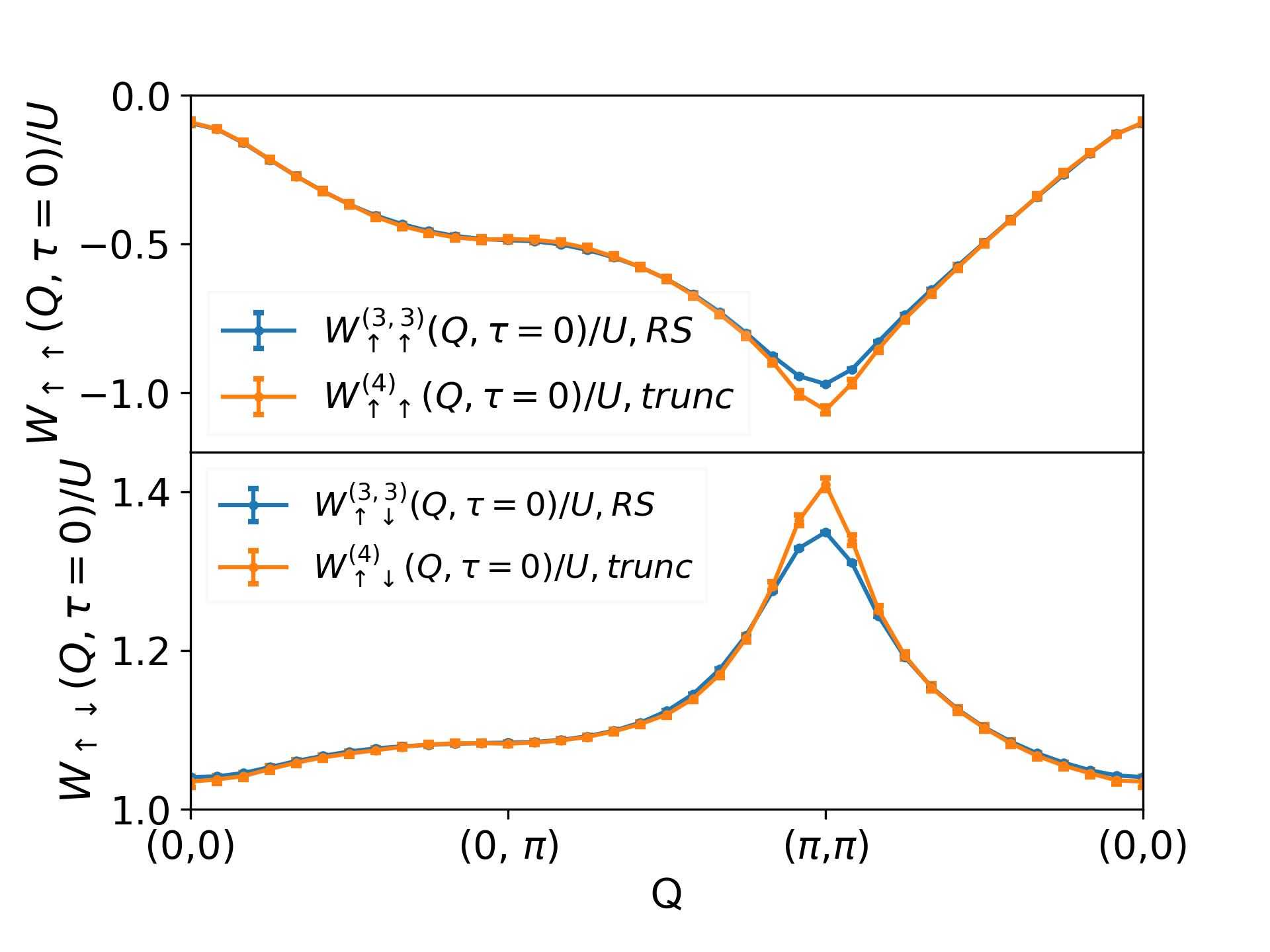}
\caption{\label{fig:QRStrunc}Resummed (RS) result for the same-time effective interaction, $W_{\sigma\sigma^\prime}^{(3, 3)}(Q, \tau = 0)$, at $\beta t= 5$ and $U/t=2$ compared to the direct expansion $W_{\sigma\sigma^\prime}^{(4)}(Q, \tau = 0)$ truncated at 4th order.}
\end{figure}

\section{Results}

\subsection{Static ($\Omega=0$) and Same-Time ($\tau=0$) Interactions}

We present the full $Q$-dependence of the effective static interaction in Fig.~\ref{fig:static} at a nominal $\beta t=3$ and $U/t=2$ where we can present the fully converged resummation of Eqs.~\eqref{eq:wuu} and \eqref{eq:wud}.  These therefore represent exact results in the thermodynamic limit.  As mentioned previously, we expected the overall sign of $W_{\uparrow\downarrow}$ and $W_{\uparrow\uparrow}$ to differ and this appears to be verified with $W_{\uparrow\uparrow}$ showing a rather substantial attractive effective interaction between same-spins.  Both curves are rather flat for much of the Brillouin zone but exhibit strong peaks near the $Q=(\pi,\pi)$ vector.  In the case of $W_{\uparrow\downarrow}$ the inclusion of higher order diagrams results in a static repulsion that is nearly double the bare $U/t$ value at $Q=(\pi,\pi)$ while the effective $W_{\uparrow\uparrow}$ becomes comparable to $U/t$ though attractive instead of repulsive.  

The actual Hubbard interaction is not a static interaction, it is a same-time interaction.  So for a fair comparison to the interaction $U$ that appears in the Hamiltonian, we compute the same-time interaction at $\tau=0$ via Eq.~\ref{eq:sametime} using AMI.  
In Fig.~\ref{fig:QRStrunc} we show an example at slightly lower temperature of $\beta t=5$ again at bare $U/t=2$.  Overall the amplitudes of the same-time case are lower than the static values which is indicative of cancellation with non-static components.  We can compute the same time objects two ways.  One is the direct  expansion of $W_{\sigma \sigma^\prime}$ including both reducible and irreducible diagrams truncated at a fixed order, and the second is the computation of $\Pi_o$ and $\Pi_e$ and using the resummations of Eqs.~\eqref{eq:wuu} and \eqref{eq:wud}.  The deviation between the resummed and truncated cases can be viewed as an uncertainty associated with the truncation of either series.  We see that at this range of $\beta$ and $U$ the result is exact except for a region near the sharp $(\pi,\pi)$ feature where the resummation somewhat softens the peak.  Nevertheless the two are in broad agreement and we use this to justify using the resummed scheme at third order.  This is important since third order calculations are substantially cheaper to compute (20 diagrams) than fourth order (137 diagrams) or the truncation at fourth order including reducible diagrams (194 diagrams in total).

\begin{figure}
\centering
\includegraphics[width=1\linewidth]{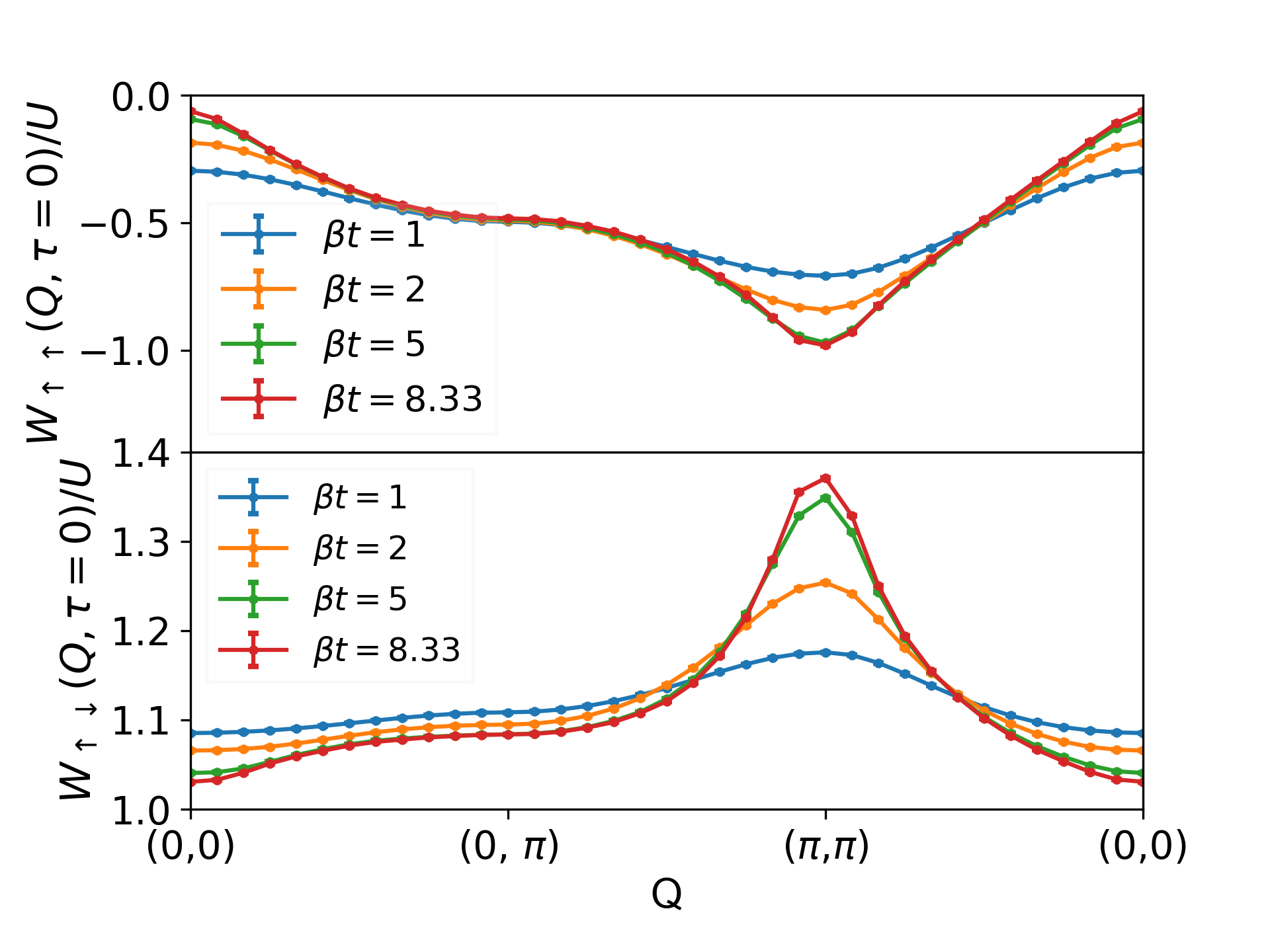}
\caption{\label{fig:Qtaubeta} The same-time effective interaction at third order, $W^{(3, 3)}(Q, \tau = 0)/ U$,  for variation in temperature and U/t = 2.}
\end{figure}

  It is well appreciated that calculations of Feynman diagrams typically become more difficult as temperature is decreased.  For the 2D Hubbard model at $\beta t=5$ and $U/t=3$ there is a metal-insulator crossover where diagrammatic methods begin to fail.\cite{fedor:2020}
Finally, we explore the temperature dependence in Fig.~\ref{fig:Qtaubeta}.
Recalling that we began with a $Q$-independent interaction such that $W_{\uparrow\uparrow}$ is zero and  $W_{\uparrow\downarrow}=U$. 
 We see that at high temperatures $\beta=1$, the effective interaction for $W_{\uparrow\downarrow}$ is only slightly above 1 but is also nearly flat in momentum.  As temperature decreases we see the emergence of the $(\pi,\pi)$ peak structure.  By $\beta t=5$ we see an overall increase in $W_{\uparrow\downarrow}$ by about 5\% at $Q=(0,0)$ increasing up to nearly 40\% near $Q=(\pi,\pi)$. Similarly, we see a strong attraction on the scale of $U/t$ between same spins.
This is a rather important difference within the model for the special $(\pi,\pi)$ nesting vector.  This result suggests that the effective interaction for that vector is substantially larger than the bare value in the Hamiltonian.  
  These results help to explain why the calculations are so much more difficult since the effective interaction is actually larger than $U/t$ by an appreciable amount and we will see that this issue worsens for larger values of the bare interaction $U/t$.

\begin{figure}
\centering
\includegraphics[width=1\linewidth]{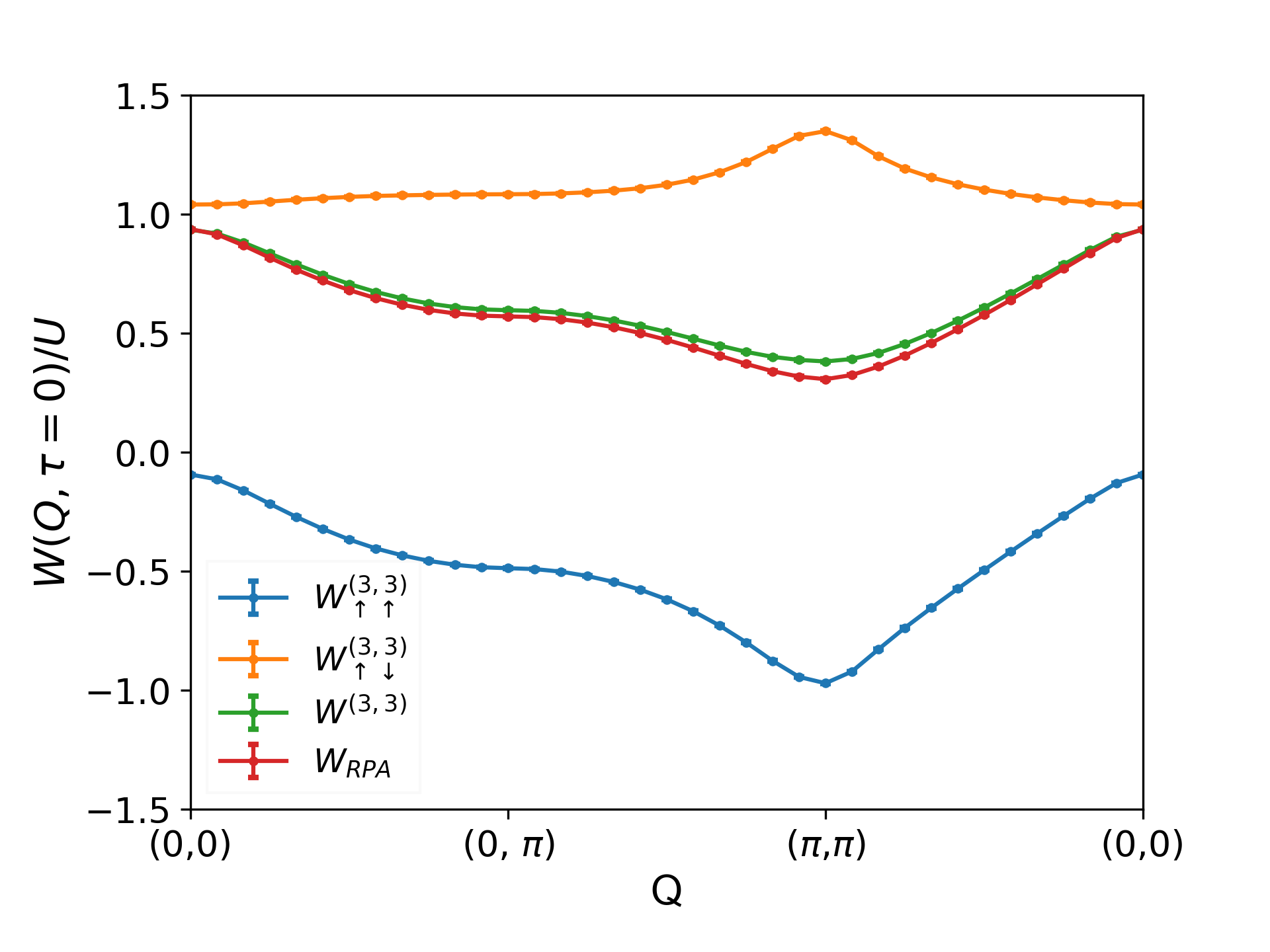}
\caption{\label{fig:Qtaurpa}Resummed result for $W^{(3,3)}_{\sigma, \sigma'}(Q, \tau = 0) / U$ and false screened interaction $W^{(3, 3)}=W^{(3,3)}_{\uparrow\uparrow}+W^{(3,3)}_{\uparrow\downarrow}$ in units of $U$ compared to the standard $W_{RPA}/U$ at $\beta t = 5$ and $U/t = 2$.}
\end{figure}

\begin{figure}
\center{\includegraphics[width=\linewidth]{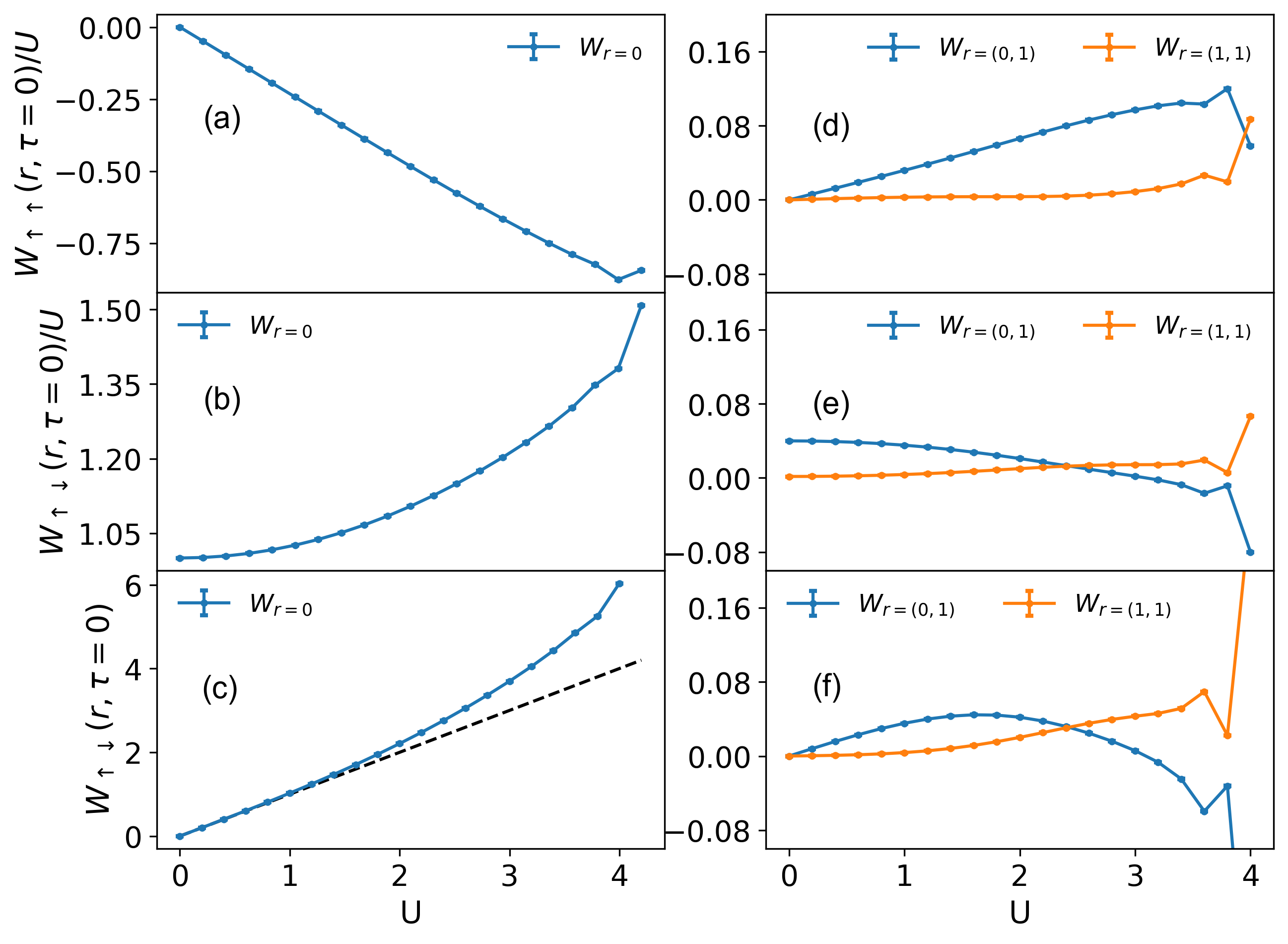}} \\
\caption{Local and non-local effective interactions $W_{\sigma\sigma'}^{(3, 3)} (r, \tau = 0) / U$ at $\beta t = 5$ as functions of $U/t$. Absolute quantities without normalization for $W_{\uparrow\downarrow}^{(3,3)}(r, \tau = 0)$ shown in frames (c) and (f).  Black dashed line in frame (c) is the unrenormalized line $y=U$ for reference. }
\label{fig:rtauVsU}
\end{figure}
\subsection{Absence of Screening}
We take an opportunity to discuss the concept of screening in interacting electron systems.  In the case of a density-density (spin independent) interaction such as the coulomb interaction we would find that all of the diagrams in $W_{\uparrow\uparrow}$ \emph{and} $W_{\uparrow\downarrow}$ represent valid renormalizations to the interaction and the total effective interaction is just the sum $W^{tot}=W_{\uparrow\uparrow} + W_{\uparrow\downarrow}$.  Shown in Fig.~\ref{fig:Qtaurpa}, we contrast the standard RPA expression based from a bare bubble to the fictitious sum of $W_{\uparrow\uparrow}$ and $W_{\uparrow\downarrow}$ from Eqs.~(\ref{eq:wuu}) and (\ref{eq:wud}).  Shown here for $U/t=2$ we are not surprised that the regular RPA expansion is a reasonable representation of the fictitious sum of the two components.  We see that there is a large cancellation between the even and odd polarization diagrams and this is the fundamental source of screening.  Hence, for coulomb-like interactions at each order of interaction there will always be matching sets of diagrams from $W_{\uparrow\uparrow}$ and $W_{\uparrow\downarrow}$ that lead to this screening.  However, this \emph{never} occurs in the Hubbard interaction.  The spin-dependent nature of the Hubbard interaction prevents mixing of these components, so while each of $W_{\uparrow\uparrow}$ and $W_{\uparrow\downarrow}$ will occur in diagrammatic expansions they will appear at different orders or in topologically distinct diagrams and will not in general trigger this cancellation.  We will see in the next section that the effective interaction becomes a runaway process with a massive repulsion of opposite spins while giving rise to an attractive interaction between same spins and these interactions can be used to infer the phases which should occur in the model.

\subsection{Local and Non-local interactions}
Since we have access to the full $Q$ dependence of the same time objects we can compute the spatial dependence of the effective interaction.  To do so we evaluate  $W_{\sigma\sigma^\prime}(Q,\tau=0)$ on a grid of size $L\times L$ in momentum for grids of $L=9,13,17,21,$ and $25$ allowing us to check that our results are relevant to the thermodynamic limit via an extrapolation in $1/L$.  This gives full control to produce accurate numerical spatial Fourier transform.  While we can do this for any $\vec{r}$, the amplitude decays sharply so we restrict discussion to the local $r=(0,0)$ as well as nearest and next-nearest $r=(0,1)$ and $r=(1,1)$ respectively.  

We fix $\beta t=5$ in Fig.~\ref{fig:rtauVsU} where we plot the $U$ dependence of the local quantities in the left hand frames (a)$\to$(c). 
We show results up to $U/t=4$ but note that the expansion is on the verge of breaking down, indicated by the erratic behaviour.  This can be rectified by including higher-orders but for what follows we focus on $U/t<4$ where the expansion remains valid. 
As expected from analytic arguments, we see that $W_{\uparrow\uparrow}/U$ is producing a linear behaviour with $U/t$ while $W_{\uparrow\downarrow}/U$ contains a clear offset of unity as well as a primarily quadratic $U^2$ behaviour.  
The behavior of $W_{\uparrow\downarrow}$ is somewhat misleading due to the offset of $1$ and $U/t$ tending to zero.  This means that the absolute $W_{\uparrow\downarrow}$, shown in Fig.~\ref{fig:rtauVsU}(c), scales linearly with $U/t$, with a predominantly $U^3$ contribution that sets in rather gently but becomes strong near $U/t=4$ at this temperature.  Similar to the case in Fig.~\ref{fig:Qtaubeta} the local quantity is enhanced by 20\% at $U/t=3$ and 50\% by $U/t=4$.  It is often the case that terms such as `weakly-coupled' or `strongly-coupled' are used as descriptors of Hubbard model systems despite the somewhat arbitrary distinction.  From our results we are motivated to suggest the distinguishing feature, that a weakly-coupled system is one where the effective interaction is comparable to (or less than) the bare interaction (the linear regime of Fig.~\ref{fig:rtauVsU}(c)) while a strongly-coupled system is one where the effective interaction is substantially larger than the bare interaction. 

Considering these results further, if one wants to understand the mechanism behind any particular phase it must be encoded in $W$.  Of particular note is the nearest-neighbor result $W_{\uparrow\downarrow}(r=(0,1),\tau=0)$ in the lower right frame of Fig.~\ref{fig:rtauVsU}.  At $U/t\approx 3$ the value switches sign from being repulsive to attractive.  These negative values begin to occur because the peak in $W(Q)$ near $(\pi,\pi)$ gains a dip and becomes two-incommensurate peaks.  The spatial dependence in the $r=(x,0)$ direction then oscillates in sign.  
This is particularly interesting in the context of the extended Hubbard model for cuprate physics where it has been suggested that including an attractive nearest-neighbour interaction promotes superconductivity\cite{jiang:2022,peng:2023} and this has been observed in 1D chain structures,\cite{chen:2021} though the latter is based in phenomenological models.
Our results suggest that even without an explicit nearest-neighbor attraction term in the Hamiltonian that non-local attraction will naturally emerge from a purely local Hubbard repulsion.  Hence, if the mechanism for superconductivity in the extended Hubbard model is non-local attraction then this might well be the mechanism for the case when the interaction is purely local. 

\begin{figure}
\centering
\includegraphics[width=1\linewidth]{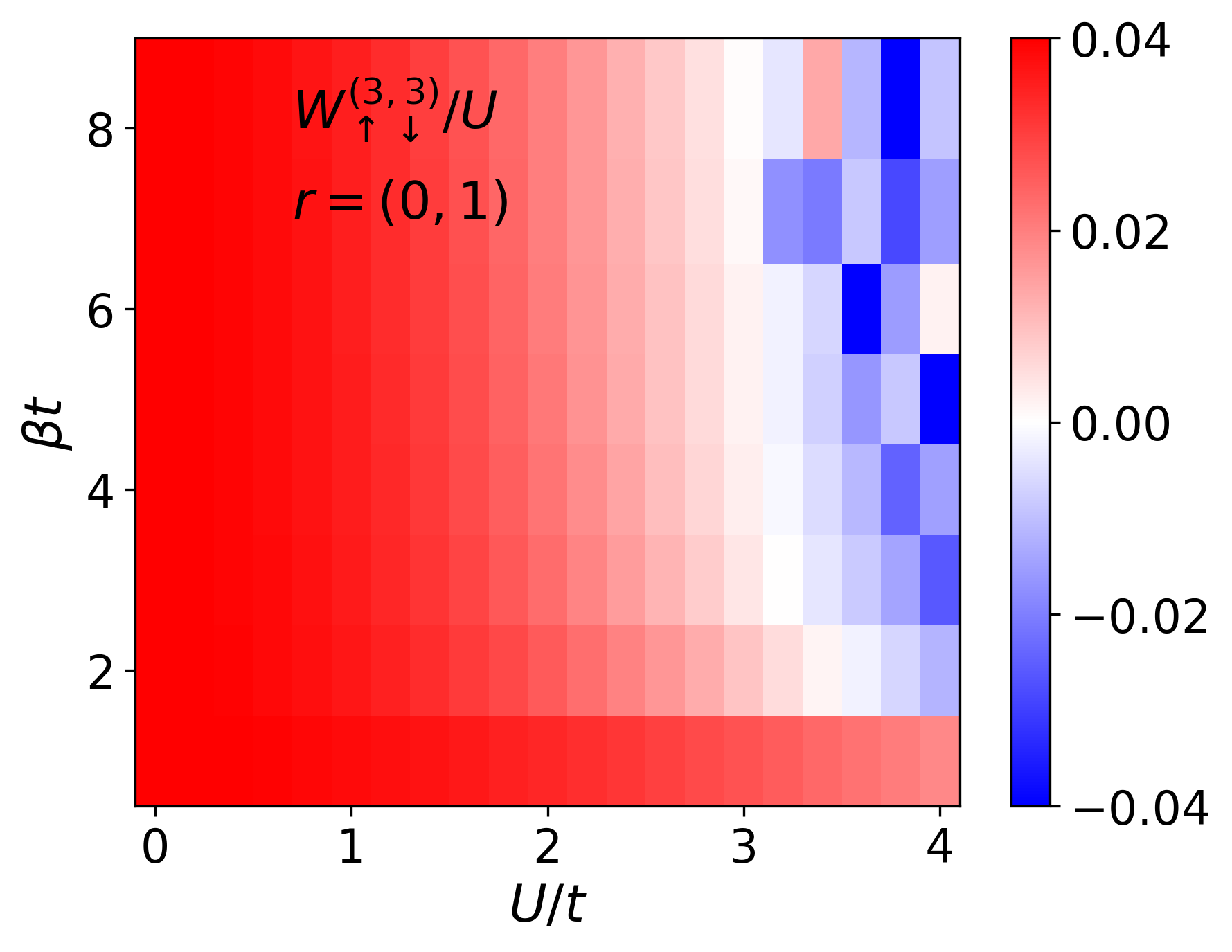}
\caption{\label{fig:countour01}False color plot of the effective nearest-neighbour $W_{\uparrow\downarrow}^{(3, 3)}(r=(0, 1), \tau = 0)/ U$ for variation in $\beta t$ and $U/t$ to identify regions of repulsive (red) and attractive (blue) behaviors.
}
\end{figure}

An obvious concern for our perturbative approach is whether the observation of non-local attraction is a robust feature of the model.  We expand upon discussion of this attraction in Fig.~\ref{fig:countour01} by plotting $W_{\uparrow\downarrow}(r=(0,1),\tau=0)$ as a function of inverse temperature and interaction strength in a false-color plot.  We see that there is a wide region in the range of $U/t=3\to4$ for $\beta t>2$ where this sign change occurs and hence there is a range of parameters where our results are controlled and reliable.  It appears to us that this range of temperature and interaction strength is similar to the regions of metal-insulator crossover and pseudogaps found in Ref.~\onlinecite{fedor:2020}.  While the concensus is that those effects are caused by $(\pi,\pi)$ spin-excitations it would seem that those effects might have an underlying imprint in the effective interaction that warrants further study.

\begin{widetext}
    
\begin{figure}
\begin{minipage}{1\linewidth}
\center{\includegraphics[width=\linewidth]{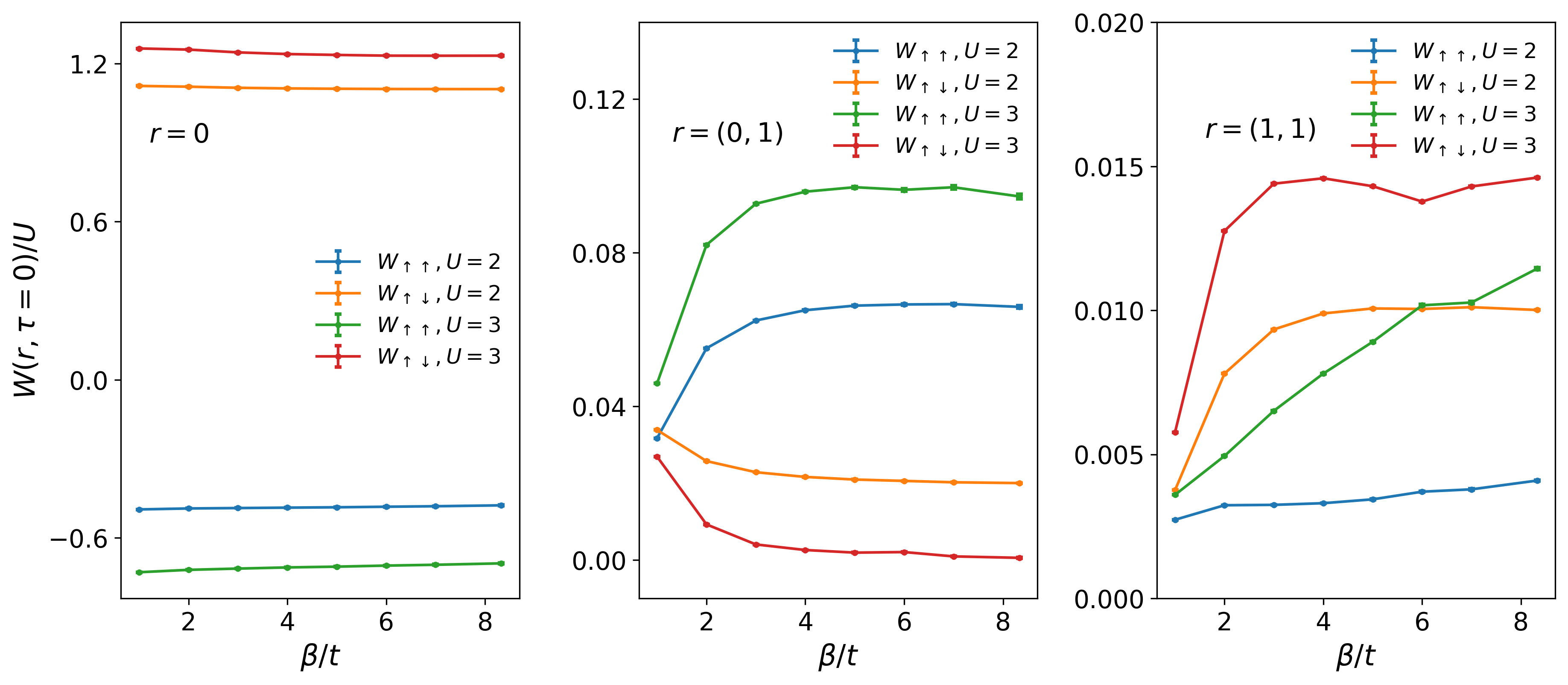}} \\
\end{minipage}
\caption{The temperature dependence of the local and non-local interactions $W^{(3, 3)}_{\sigma \sigma'}(r, \tau = 0) / U $.}
\label{fig:rtauvsbeta}
\end{figure}

\end{widetext}

These spatial correlations set in as temperature is decrease.  In Fig.~\ref{fig:rtauvsbeta} we plot fixed $U/t$ slices of Fig.~\ref{fig:countour01} at values of $U/t=2$ and $3$, as well as the local and second-nearest neighbour equivalents, to illustrate the dependence on inverse-temperature $\beta$.  The first key insight is that the local effective interaction does not strongly depend on temperature, and while it does depend on the value of $U/t$ we find extremely flat temperature dependence over this range. Instead, the effects of temperature are seen starkly in the effective non-local interactions.  It seems that whatever physical processes are occurring that the local physics is somewhat frozen while the non-local is very dynamic with temperature.  This has catastrophic consequences for many numerical embedding methods such as dynamical mean-field theory based around the solution of a local Anderson impurity.\cite{footprints,leblanc:2013}  It suggests that taking only local physics will entirely miss the temperature dependent features of the effective interaction that seems to be responsible for antiferromagnetism as well as providing an attractive channel for pairing.  The propensity for antiferromagnetism is directly apparent in the effective-interaction.  For example, one can see that in the $r=(0,1)$ or $(1,0)$ directions $W_{\uparrow\uparrow}$ is repulsive and growing as $\beta$ increases while $W_{\uparrow\downarrow}$ is decreasing and always lower amplitude than $W_{\uparrow\uparrow}$.  It is therefore becoming energetically favourable to have an antiferromagnetic configuration.
Similarly in the diagonal $r=(1,1)$ direction the situation is reversed with $W_{\uparrow\uparrow}<W_{\uparrow\downarrow}$, which again makes it favourable to orient same-spins along the diagonals and again pushing the system to be antiferromagnetic.

\section{Conclusions}

The details of the fully renormalized interactions for correlated electron systems can provide a qualitative understanding of the phases present in a model.  In the case explored here, the 2D square lattice model, by starting with only a local same-time repulsion between opposite spins, the effective renormalized interaction becomes larger than the bare value of $U/t$.  Unlike typical density-density interactions the effective interaction in the Hubbard model does not exhibit screening processes.  In particular it is peaked near $Q=(\pi,\pi)$ for the half-filled model and the resulting local same-time object can be substantially enhanced from the bare value by as much as $40\%$ in our explored parameter range.  We find two cases for attractive interactions: 1) the effective same-spin interaction is attractive for all momenta, and 2) we find the emergence of an attractive nearest-neighbour interaction between opposite spins along the nearest-neighbor, $r=(0,1)$, direction.

That the Hubbard Hamiltonian produces a non-local attraction is perhaps not surprising given the vast literature observing superconductivity on finite-sized 2D square lattices at finite temperatures.\cite{gull:2015,gull:pnas:2022}  Any time an attractive interaction exists it is expected that pairing can occur on some length scale.  Our results show specifically that short range attraction between opposite spins exists as a property of the weakly-coupled Hubbard model.  It is therefore reasonable to suggest that this is a dominant pairing mechanism as observed in the extended Hubbard model.\cite{peng:2023}  What is not yet understood is the role of stripe phases, and if their existence prevents a macroscopic superconducting ground state.

\bibliographystyle{apsrev4-2}
\bibliography{refs.bib}

\end{document}